\documentclass[runningheads]{llncs}
\usepackage[T1]{fontenc}
\usepackage{multirow}
\usepackage{bbding}
\usepackage{pifont}
\usepackage{subfigure}
\usepackage{caption}
\usepackage{graphicx}
\usepackage{tabularray}
\usepackage{booktabs}
\usepackage{makecell} 

\usepackage{algorithm}
\usepackage{algorithmic}
\usepackage{subcaption}
\usepackage{enumitem}

\usepackage{amsmath}

\newcommand{\minisection}[1]{\vspace{5pt}\noindent\textbf{#1.}}

\usepackage{graphicx}
\begin{document}
\title{Retrieval and Distill: A Temporal Data Shift-Free Paradigm for Online Recommendation System}
\titlerunning{Retrieval and Distill}
%
\author{Lei Zheng\inst{1} \and
Ning Li\inst{1}\and
Chengguang Gan\inst{2} \and
Yong Yu \inst{1} \and
Weinan Zhang \inst{1}
}
\institute{
Shanghai Jiao Tong University, Shanghai, China\\
\email{\{zhenglei2016,lining01,yyu,wnzhang\}@sjtu.edu.cn} \and
Yokohama National University, Yokohama, Japan\\
\email{ganchnegguan@yahoo.co.jp}
}

%
\maketitle              
\begin{abstract}

Current recommendation systems are significantly affected by a serious issue of temporal data shift, which is the inconsistency between the distribution of historical data and that of online data.
Most existing models focus on utilizing updated data, overlooking the transferable, temporal data shift-free information that can be learned from shifting data.
We propose the Temporal Invariance of Association theorem, which suggests that given a fixed search space, the relationship between the data and the data in the search space keeps invariant over time.
Leveraging this principle, we designed a retrieval-based recommendation system framework that can train a data shift-free relevance network using shifting data, significantly enhancing the predictive performance of the original model in the recommendation system.
However, retrieval-based recommendation models face substantial inference time costs when deployed online.
To address this, we further designed a distill framework that can distill information from the relevance network into a parameterized module using shifting data.
The distilled model can be deployed online alongside the original model, with only a minimal increase in inference time.
Extensive experiments on multiple real datasets demonstrate that our framework significantly improves the performance of the original model by utilizing shifting data. 
\keywords{Temporal Data Shift, Retrieval-Enhanced Methods, Knowledge Distillation, Pretraining, Click-Through Rate}
\end{abstract}

\section{INRODUCTION}

Click-through rate (CTR) prediction models play a significant role in modern recommendation systems. To enhance the accuracy of CTR models, deep learning-based approaches have been extensively employed \cite{covington2016deep}\cite{shan2016deep}\cite{wang2017deep}\cite{wang2021dcn}. However, the dynamic environment faced by the recommendation systems, with the data distribution evolving over time, poses significant challenges to the effectiveness of these models \cite{zhu2023reloop2}.

To tackle this challenge, prior research has undertaken investigations from two distinct perspectives, namely incremental learning and behavior sequence modeling. 
On one hand, incremental learning techniques delve into updating the models that make these predictions as new data comes in, without having to start from scratch every time \cite{ye2022future}\cite{peng2021learning}\cite{katsileros2022incremental}. 
This means the models learn and improve gradually as they get access to more recent data. 
However, the process of learning neural network parameters depends on iterative updates based on gradients, which slowly infuse supervision data into the model weights using a minimal learning rate. 
This poses a significant challenge for large parametric recommendation models in swiftly adapting to shifts in distribution.
On the other hand, behavior sequence modeling focuses on understanding how users' behaviors change over time by looking at the sequence of actions they take. 
In this area, various network structures have been proposed, including recurrent
neural networks (RNNs) \cite{hidasi2015session}\cite{hidasi2016parallel}, convolutional neural networks (CNNs) \cite{tang2018personalized}\cite{yuan2019simple}, and memory networks \cite{pi2019practice}\cite{ren2019lifelong}.
Moreover, the attention mechanism has emerged as the predominant approach for modeling item dependencies, with several notable contributions such as SASRec \cite{kang2018self}, DIN \cite{zhou2018deep}, DIEN \cite{zhou2019deep}, and BERT4Rec \cite{sun2019bert4rec}.
These studies aim to identify sequential patterns in user behavior, thereby facilitating the prediction of future preferences based on historical activities. However, these studies fail to directly tackle the issue of distribution shift during training time, and these high-capacity models significantly increase the online workload and inference burden.

\begin{figure}[t]
\centering 
\includegraphics[width=0.50\textwidth]{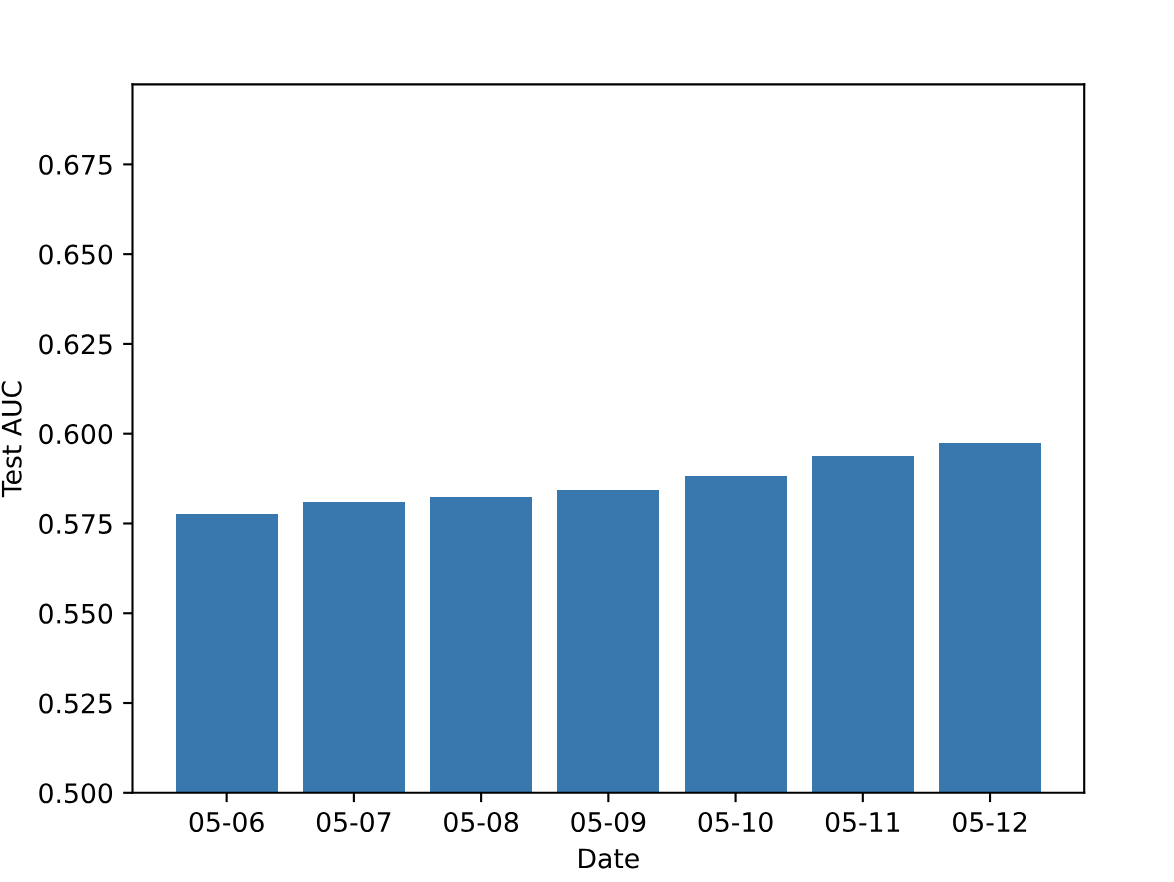} 
\vspace{-10pt}
\caption{The phenomenon of temporal data shift on a real dataset collected from Tmall. The test set consists of data after Nov. 10, 2015. The x-axis represents the date from which we add all data up to Nov. 10 to train a CTR model, and the y-axis represents the model's performance on the test set. 
We observed that model performance peaked when trained on temporally adjacent data, with performance degrading as the training period extended further from the test set timeframe.
}
\label{fig:tmallshift}
\vspace{-12pt}
\end{figure}

All of the above-mentioned models either update the model using the latest data \cite{guo2017deepfm,wang2017deep} or use past shifting data as a reference for retrieval \cite{qin2020user,qi2020search}, without directly using the shifting data for training model parameters.
However, directly using the temporally shifting data for model training is challenging, as an example illustrated in Figure~\ref{fig:tmallshift} based on a public dataset of Tmall, an international e-commerce platform.
In this dataset, data spanning Nov. 11-12, 2015 is employed as the test set.
The x-axis represents the training set from the indicated date to Nov. 10, 2015. We train a common deep CTR model PNN \cite{qu2018product} using different training sets to make predictions. The y-axis indicates the AUC performance of the model on the test set.
We can observe that utilizing data over Nov. 1-10, 2015 for predictions on the test set yields the best results, whereas adding data from earlier time periods will negatively impact the results due to data shift.
Therefore, it is crucial to find an effective method that can directly utilize these data for model training.

Inspired by the recent proposed retrieval-based tabular data prediction models \cite{qin2021retrieval,zheng2023dense}, we found that similar data from historical records can significantly enhance the predictive capability and generalizability.
In other words, although the probability distribution modeled by the CTR is time-variant, the probability distribution of the association between current user behavior data and similar historical data can remain invariant over time if the search space is fixed.
If we design a neural network to model the probability distribution of data similar to the current query, this neural network is temporal data shift-free. Thus, the shifting data can be used to train such a neural network.
By incorporating this neural network with the original CTR model, we can obtain an aggregated prediction model that has been trained with shifting data.

In this paper, we propose the Retrieval and Distill paradigm (RAD) to address the challenge of ineffective training of a CTR model over shifting data.
The RAD paradigm generally consists of two major components, namely the Retrieval Framework and the Distill Framework. 
We first establish a Retrieval Framework to assist the original model in making predictions using similar data within the shifting data. 
We refer to the neural network in Retrieval Framework that retrieves and outputs the logits of similar data as the relevance network. 
By pre-training this relevance network using shifting data and then merging the original CTR model and relevance network, we can significantly enhance the predictive capability of the original CTR model. 
However, the added relevance network in the Retrieval Framework, due to its retrieval process (such as BM25 \cite{robertson1995okapi}), is difficult to deploy online due to its inference time, similar to other retrieval-based CTR models \cite{qin2021retrieval}\cite{zheng2023dense}. 
To address this difficulty, we further build the Distill Framework to distill knowledge from the relevance network using shifting data, obtaining a parameterized student network called search-distill module. 
By aggregating the original CTR model and the search-distill module, we finetune the Distill Framework with unshifting data, e.g., the latest training data.
This yields the final model that can be deployed online for light workload and efficient inference.


In sum, the technical contributions of this paper are threefold.
\begin{itemize}[leftmargin=10pt]
    \item We highlight the significant impact of the data shift phenomenon in recommendation systems, which is not addressable by simply increasing training data or the model capacity. To deal with this issue, we propose the Retrieval and Distill (RAD) paradigm, where the recommendation model is enhanced by incorporating shifting data into the model. To the best of our knowledge, RAD is the first paradigm that successfully utilizes shifting data for model training and lightweight deployment.
    \item As the rationale of RAD, we raise the theorem of Temporal Invariance of Association, which provides effective guidance on how to utilize shifting data.
    \item We design an innovative distillation approach that distills information from the retrieval framework of RAD. This approach optimizes both the time complexity and space complexity of the retrieval process to $O(1)$, which largely pushes forward the practical usage of retrieval-based models for online recommendation systems. 
\end{itemize}


Extensive experiments on multiple real datasets demonstrate that RAD significantly improves the performance of the original model via effective training over shifting data.
RAD offers a new paradigm in the field of recommendation systems, enabling models to train using both shifting and unshifting data, opening up a new pathway for enhancing the performance of CTR models by increasing the amount of training data.

\section{PRELIMINARIES}\label{sec:preliminaries}

In this section, we formulate the problem and introduce the notations. 
In the context of recommendation systems, we represent the dataset as a tabular structure, denoted by \( \mathcal{D} \), comprising \( F \) feature columns \( \mathcal{C} = \{C_i\}_{i=1}^F \) alongside a singular label column \( \mathcal{Y} \). 
The feature columns encapsulate attributes such as "Age," "City," and "Category."
Each individual entry within table \( \mathcal{D} \) corresponds to a sample \( s_z \). 
A sample is characterized by an amalgamation of multiple features along with a label, articulated as \( s_z = (x_z, y_z) \), where \( x_z = \{c_i^z\}_{i=1}^F \). 
Accordingly, table \( \mathcal{D} \) is defined as an ensemble of \( N \) samples, expressed as \( \mathcal{D} = \{s_z\}_{z=1}^N \).

A click-through rate (CTR) model can be mathematically represented as a conditional probability, $P(y|x)$, where $y$ is the event of a click, and $x$ represents the features. The goal of the model is to accurately estimate this probability given the features $x$, which can include information about the item, the user, and the context.

\begin{theorem}\label{theorem:xs}
Temporal Invariance of Association.
\end{theorem}

In the paper, we model $P(y|x)$ as $P(y|x, x_s)$, where $x_s$ represents documents similar to $x$ within a fixed search space $\mathcal{E}$.
Given that the probability distribution of $P(y|x, x_s)$ changes over time, while the probability distribution of $P(x_s|x)$ remains constant, we can utilize the data come from the time-varying $P(y|x, x_s)$ to train a model that represents $P(x_s|x)$. 
This process allows us to capture the temporal dynamics in $P(y|x)$ while leveraging the time-invariant properties of $P(x_s|x)$, thereby enhancing the model's adaptability to temporal changes in the future.

We partition the dataset by date into three segments. 
The last few days, denoted as \( \mathcal{D}_{[t+1, t+\Delta]} \), serve as the test set, \( \mathcal{D}_{\text{test}} \). 
Data within a window of \( d \) days prior to the test dates is considered unshifting data, which is \( \mathcal{D}_{[t - d + 1, t]} \) and is used as the training set, denoted by \( \mathcal{D}_{\text{train}} \). 
The remaining portion, \( \mathcal{D}_{(-\infty, t - d]} \), is designated as the search space, \( \mathcal{D}_{\text{shifting}} \).

The \textbf{Retrieval and Distill (RAD)} framework comprises two components. 
The first component, the Retrieval Framework, employs a retrieval-based method to assist any original model in addressing the data shift issue. 
However, retrieval-based models face significant efficiency challenges. 
Therefore, the second component, the Distill Framework, utilizes distillation to eliminate the retrieval process. 
We will introduce these two components separately in the following.

\minisection{Retrieval Framework}
The objective of the Retrieval Framework is to employ a relevance network, which incorporates non-parametric retrieval, to assist any original model $F(x)$ in predicting the label. The formulation of the Retrieval Framework is as follows:
\begin{equation}
    \hat{y_R} = f_\theta(F(x), x, R(x)), \label{eq:retrieval_model}
\end{equation}
where $R(x)$ is designed to fit $P(x_s|x)$ and is referred to as the \textit{relevance network} and $f_\theta$ which approximate the $P(y|x, x_s)$ is data shift sensitive.

To obtain a model, $R(x)$, that approximates the probability distribution $P(x|x_s)$, we employ a Retrieval Framework, devoid of $F(x)$, named the teacher network. The formulation of the teacher network is as follows:
\begin{equation}
    \hat{y_T} = T_\gamma(x, R(x)). \label{eq:retrieval_model}
\end{equation}

According to Theorem~\ref{theorem:xs}, The $R(x)$ component within $T_\gamma$, trained using shifting data, is invariant to data shift.
Therefore, to obtain an $R(x)$ that has been trained with shifting data, $T_\gamma$ can be minimized by the following loss function:
\begin{equation}\label{loss:retrieval_pretrain}
   \mathcal{L}_{\text{pretrain}} =  \sum_{x_z \in \mathcal{D}_{\text{shifting}}} \text{CE}(y^z, \hat{y}_T^z),
\end{equation}
where $CE$ is binary cross-entropy. This step obtains the model $R(x)$ that approximates $P(x_s|x)$.

Once we have obtained an $R(x)$ that has been pretrained with shifting data, we can utilize $R(x)$ in the training of the Retrieval Framework. We use unshifting data to minimize the following loss function:
\begin{equation}\label{loss:retrieval_finetune}
   \mathcal{L}_{\text{retrieval}} =  \sum_{x_z \in \mathcal{D}_{\text{train}}} \text{CE}(y^z, \hat{y}_R^z),
\end{equation}
where $\text{CE}$ is binary cross-entropy. In this step, we obtain the Retrieval Framework, $f_\theta$, which approximates $P(y|x, x_s)$.

\minisection{Distill Framework}
Although the Retrieval Framework can leverage shifting data to significantly enhance the original model, the deployment of retrieval-based methods presents substantial challenges in terms of storage space and inference time in an online environment. Therefore, we have designed the Distill Framework to eliminate the non-parametric search process.

Distill Framework can be formulated as:
\begin{equation}
    \hat{y}_D = f'_\phi(F(x), x, r(x)),\label{eq:retrieval_model}
\end{equation}
where $r(x)$ is distilled from $R(x)$ pretrained using loss function ~(\ref{loss:retrieval_pretrain}), which we refer to as the \textit{search-distill module}. Similar to $f_\theta$, $f'_\phi$ approximates $P(y|x, x_s)$ and can also only be trained using unshifting data.

We assume
\begin{equation}
    w_R = R(x),
\end{equation}
where $w_R$ is the last layer logits of $R(x)$ and assume
\begin{equation}
    w_r = r(x),
\end{equation}
where $w_s$ is the last layer logits of $r(x)$.
The loss function used to distill knowledge from $R(x)$ to $r(x)$ can be written as:
\begin{equation}\label{loss:kd}
   \mathcal{L}_{\text{KD}} =  \sum_{x_z \in \mathcal{D}_{\text{shifting}}} \text{MSE}(r(x), R(x))
= \sum_{x_z \in \mathcal{D}_{\text{shifting}}} \text{MSE}(w_r^z, w_R^z),
\end{equation}
where $MSE$ is the Mean Squared Error loss function.
Once we obtain the search-distill module, we can use unshifting data to derive a final Distill Framework using the following loss function:
\begin{equation}\label{loss:distill_finetune}
   \mathcal{L}_{\text{distill}} =  \sum_{x_z \in \mathcal{D}_{\text{train}}} \text{CE}(y^z, \hat{y}_D^z),
\end{equation}
where $CE$ is binary cross-entropy loss.

\section{The RAD Paradigm}\label{sec:framework}

In this section, the details of RAD will be presented. We start with an overview of the entire paradigm's structure at Section~\ref{sec:overview}, followed by detailed expositions of the Retrieval Framework and Distill Framework in Sections~\ref{sec:retrievalframework} and~\ref{sec:distillframework}, respectively.

\begin{figure*}[t]
\centering 
\includegraphics[width=0.7\textwidth]{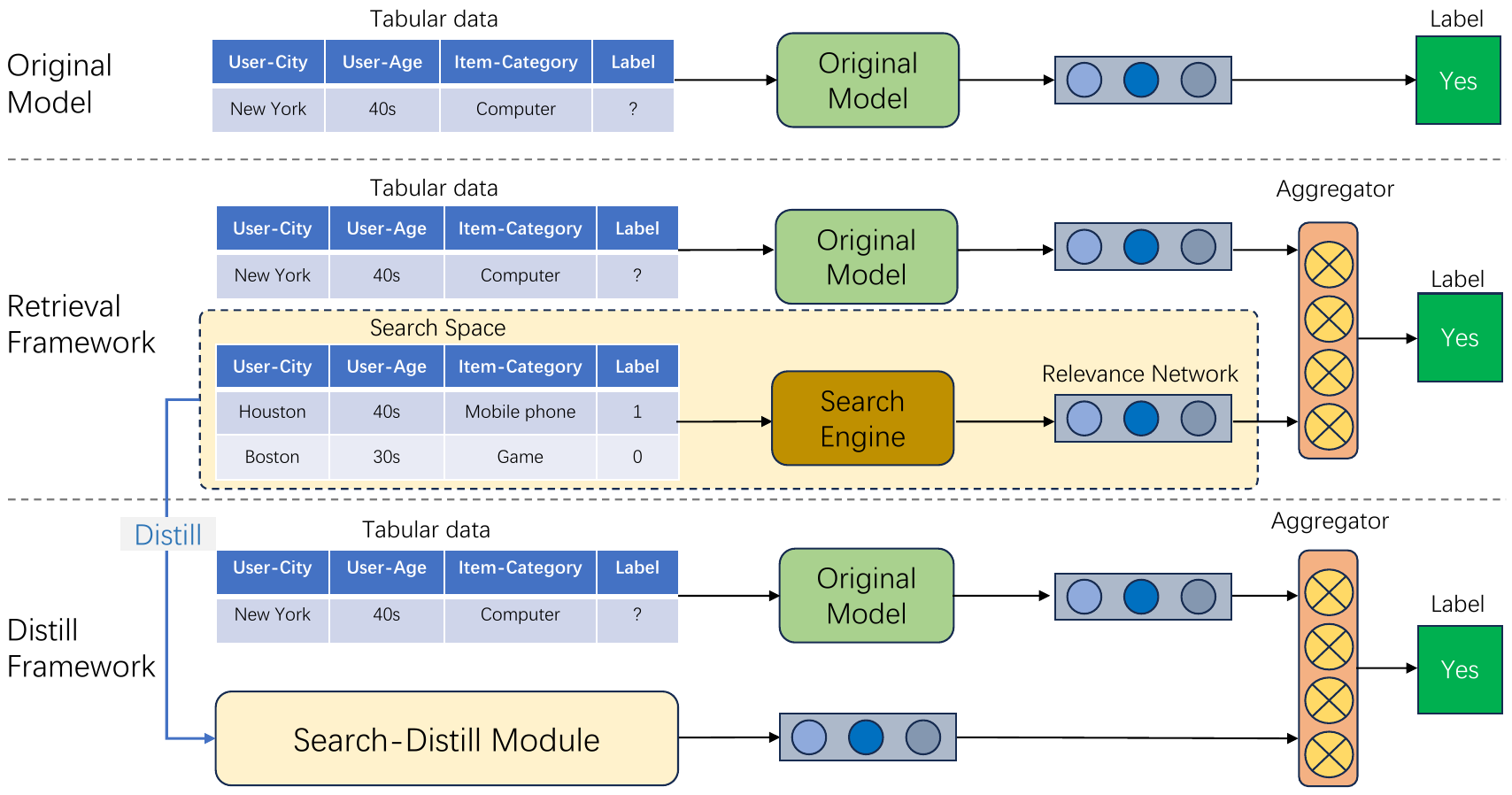} 
\caption{This figure illustrates the original model and its combination with the two components of the RAD paradigm. In CTR tasks, the original model typically takes tabular data as input and outputs a probability indicating the likelihood of a click. The RAD paradigm initially employs a retrieval framework to assist the original model in making predictions using shifting data. However, considering the performance requirements of the CTR model, we designed the RAD's distill framework to eliminate the search process in the retrieval framework, thereby meeting the online inference time requirements.}
\label{fig:overview}
\vspace{-10pt}
\end{figure*}

\begin{figure}[t]
\centering 
\includegraphics[width=0.6\textwidth]{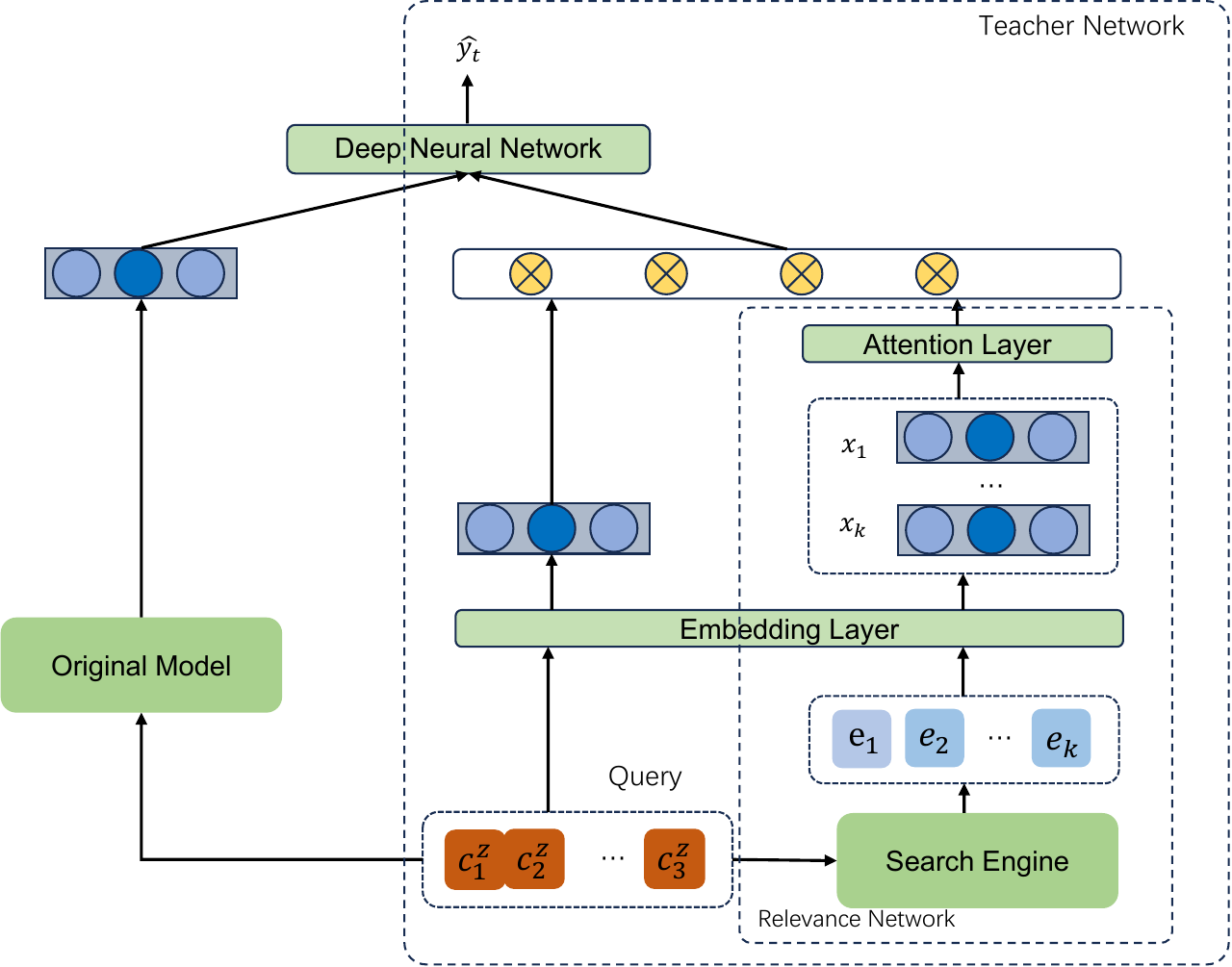} 
\vspace{-10pt}
\caption{Retrieval framework.}
\label{fig:retrievalframe}
\vspace{-10pt}
\end{figure}

\begin{figure}[t]
\centering 
\includegraphics[width=0.7\linewidth]{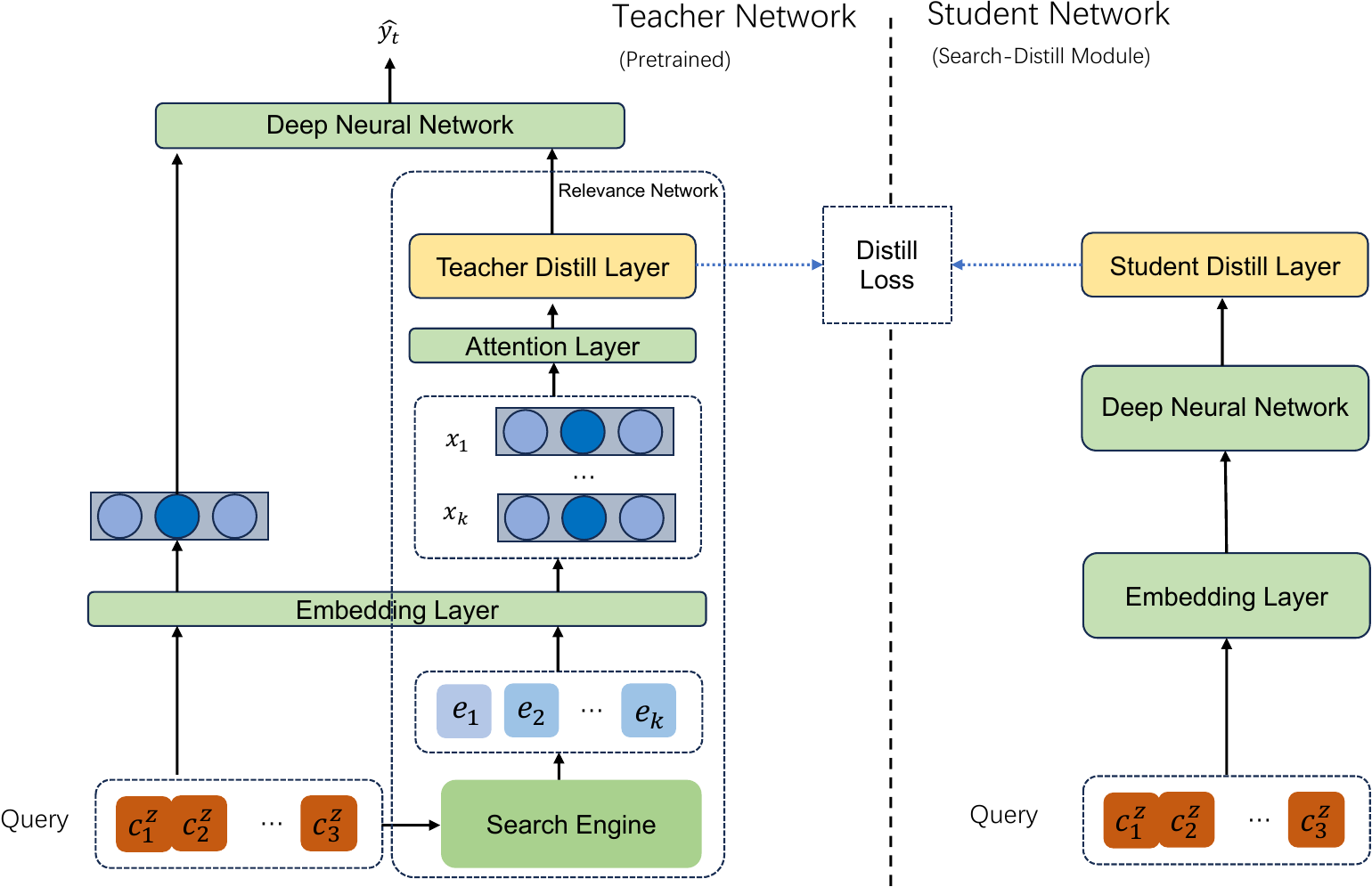} 
\vspace{-5pt}
\caption{Distill Process.}
\label{fig:distillprocess}
\vspace{-20pt}
\end{figure}

\begin{figure}[t]
\centering 
\includegraphics[width=0.55\textwidth]{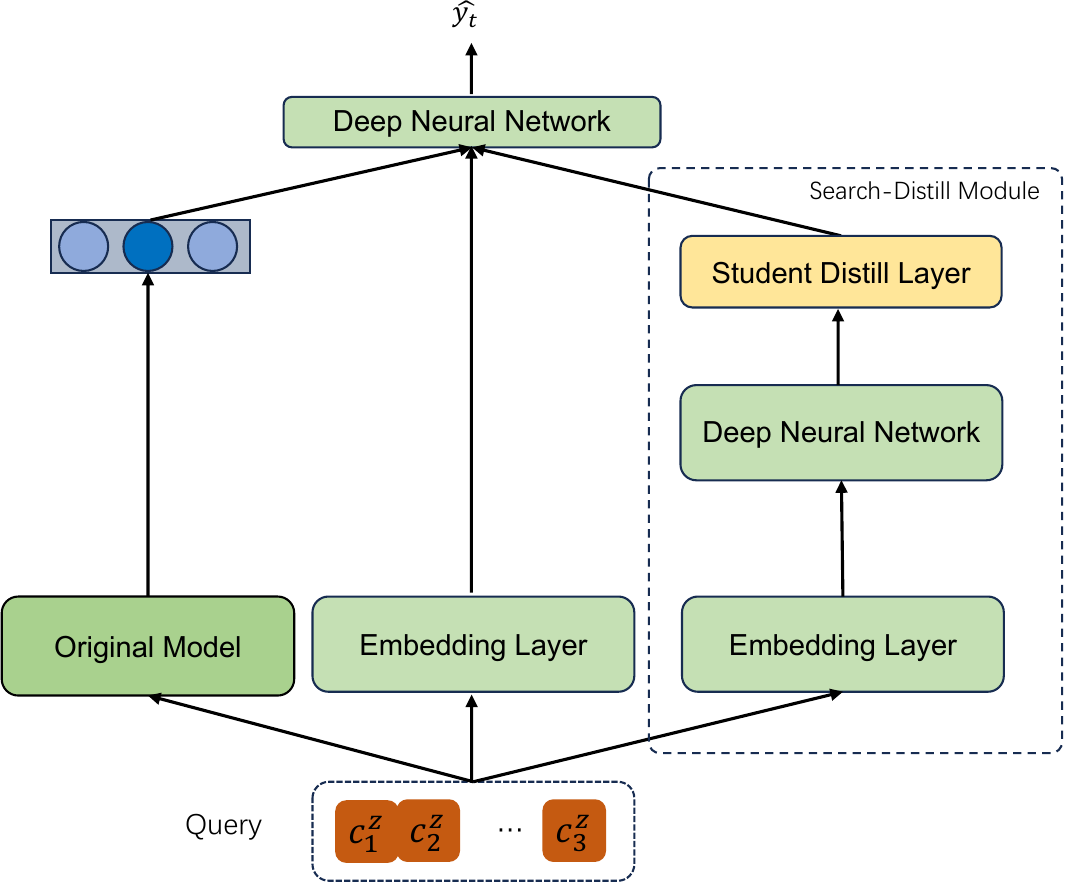} 
\caption{Distill Framework.}
\label{fig:distillframework}
\vspace{-15pt}
\end{figure}

\subsection{Overview}\label{sec:overview}
Traditional CTR models have not accounted for the data shift issue, and they cannot enhance performance by merely adding shifting data \cite{juan2016field}\cite{naumov2019deep}\cite{gu2020deep}. 
Training only with unshifting data neglects a significant amount of information.
Our proposed Retrieval and Distill paradigm aims to devise a universal solution that utilizes shifting data to enhance the performance of existing models. 
RAD is generally divided into two parts.
The first part incorporates a Retrieval Framework to introduce shifting data into the datastore during training. 
Concurrently, as per theorem \ref{theorem:xs}, we pretrain the model's relevance network with shifting data and then transfer it for use. 
This approach substantially improves the model's performance without altering its original structure. 
However, including a non-parametric search component in the relevance network poses deployment challenges for online recommendation systems. 
Therefore, the second part is a Distill Framework tasked with transforming the relevance network into a parametric neural network model. 
We distill a simple neural network (search-distill module) using shifting data and the relevance network as the teacher network. 
Ultimately, we merge the search-distill module with the original model and finetune it using unshifting data to yield the final online deployable model. 
The following sections will elucidate these modules in detail.

\vspace{-10pt}
\subsection{Retreival Framework} \label{sec:retrievalframework}
The Retrieval Framework aims to employ a relevance network that utilizes shifting data in conjunction with the original model for CTR prediction.
According to theorem~\ref{theorem:xs}, the relevance network is temporal data shift-free; therefore, as an initial step, we pretrain the relevance network using shifting data with a teacher network, which is a Retrieval Framework that does not contain the original model. In the second step, we jointly train the relevance network and original model using unshifting data.
Next, we will detail the neural network structure of the Retrieval Framework,  followed by its training method.

\subsubsection{Framework Model Structure}\label{sec:retrieval-structure}
The neural network structure of the Retrieval Framework can be seen in Figure~\ref{fig:retrievalframe}.
The Retrieval Framework aggregates the outputs of both the teacher network and the original model, where the teacher network specifically combines the original input with the relevance network. We shall now explicate the original model, relevance network, and teacher network.

\minisection{Original Model} We denote \( s_z \) as any tabular data entry within the dataset \( \mathcal{D} \), where the feature part of \( s_z \) is represented as \( x_z \), and the label part as \( y_z \), with \( x_z = \{c_z^i\}_{i=1}^F \) comprising \( F \) features. Traditional CTR models take such an \( x_z \) as input and output a probability.
We represent the original model as \( O \), with the last layer logit denoted by \( w_o \). 

In the traditional model, the probability prediction is given by 
\begin{equation}
    \hat{y_O^z} = O(x_z) = \text{sigmoid}(w_o),
\end{equation}
where $w_o$ is the input of the aggregation layer of the Retrieval Framework.

\minisection{Relevance Network}
The relevance network features an embedding layer shared with input $x_z$, a search engine, and an attention-based aggregation layer.
Within our framework, we utilize shifting tabular data entries as the search space, with BM25 \cite{robertson1995okapi} serving as our search algorithm.

In the BM25 algorithm, we treat any sample $x_z$ as the query and a key $x_d$ in the datastore as the document. As such, the ranking score can be calculated as
\begin{align}
    \text{RankScore}(x_z, x_d) &= \sum_{j=1}^M \text{IDF}(a_j^z)\cdot \mathbf{1} (x_j^z = x_j^d) ,\\
    \text{IDF}(x_j^z) &= \log\frac{N-N(a_j^z) + 0.5}{N(a_j^z) + 0.5},
\end{align}
where $\mathbf{1}(\cdot)$ is the indicator function, $N$ is the number of data samples in $\mathcal{D}_{\text{shifting}}$, and $N(a_j^z)$ is the number of data samples that have the categorical feature value $x_j^z$. By ranking the candidate samples using BM25, we obtain the
top $K$ entries as the retrieved set $\mathcal{S}(x_z)$.

For each item within \( \mathcal{S}(x_z) \), an attention mechanism is applied to calculate its aggregation parameter
\begin{equation}
    \alpha_k = \frac{\exp(x_k^\top W x_z)}{\sum_{j=1}^{K} \exp(x_j^\top W x_z)},
\end{equation}

The elements of \( \mathcal{S}(x_z) \) are then consolidated into a  logit \( w_R \) : 
\begin{equation}
    w_R = \sum_{k=1}^{K} \alpha_k \cdot x_k, \quad s_k \in \mathcal{S}(x_z),
\end{equation}
where $w_s$ is the input of the aggregation layer of the teacher network.

\minisection{Teacher Network} The teacher network is a Retrieval Framework that does not include the original model. From Figure~\ref{fig:retrievalframe}, we can observe that the teacher network forms the right portion of the diagram.
The teacher network serves as an intermediary bridge for training the relevance network.
The teacher network takes $x_z$ and the relevance network output as inputs, producing a probability as output.
Drawing inspiration from RIM \cite{qin2021retrieval}, we incorporate a factorization machine module within the aggregation layer of the teacher network. 
This aggregation layer is designed to effectively merge the input $x_z$  with the aggregated $w_R$, thus improving the teacher network's capability to discover the hidden relationships between $x_z$ and the returned $S(x_z)$ from the search engine.

\subsubsection{Pretrain with Shifting Data}\label{sec:retrieval-pretrain}
The relevance network models the probability distribution \( P(w_R|x) \), where \( w_R \) is the logit representation of \( x_s \). As established by theorem~\ref{theorem:xs}, the probability distribution represented by the relevance network is time-invariant, which allows us to pretrain it using shifting data. 
To obtain the relevance network, we initialize a separate teacher network. The teacher network is trained using the dataset \( \mathcal{D}_{\text{shifting}} \) and with loss function ~(\ref{loss:retrieval_pretrain}).
Although the separate teacher network trained in this manner performs poorly on the test set \( \mathcal{D}_{\text{test}} \) due to data shift—often hovering around 0.5 in our experiments—the relevance network is transferable.

\subsubsection{Finetune with Unshifting Data}\label{retrieval-finetune}

After obtaining the pretrained relevance network, we integrate it with the original model and the target query \( x_z \) to form the Retrieval Framework. This framework is then trained using \( \mathcal{D}_{\text{train}} \) with loss function ~(\ref{loss:retrieval_finetune}), essentially finetuning it. 

\subsection{Distill Framework}\label{sec:distillframework}


To optimize the time complexity of the Retrieval Framework, we propose a Distill Framework that distills the relevance network (pretrained via the Retrieval Framework) into a search-distill module by removing its computationally intensive BM25 search component. This module is then integrated with the original model through a multilayer neural network, aggregated, and finetuned using \( \mathcal{D}_{\text{training}} \), enabling efficient deployment in online environments.

\subsubsection{Framework Model Structure}\label{sec:distill-structure}
The neural network structure of the Retrieval Framework can be seen in Figure~\ref{fig:distillframework}.
The network architecture of the Distill Framework is similar to that of the Retrieval Framework. We replace the relevance network with the search-distill module. 

From Figure~\ref{fig:distillprocess}, we can observe the form of the teacher network in the Distill Framework.
The teacher network in the Distill Framework differs somewhat from that in the Retrieval Framework. We removed the cross layer from the teacher network in the Retrieval Framework to distill only the relevance network portion and added two neural network layers after the attention layer. 
We use this teacher network to obtain a relevance network for distillation.
 
The search-distill module is a multilayer neural network equipped with an embedding layer. Akin to the original model, the search-distill module takes the feature portion of tabular data $x_z$ as its input. The output of the search-distill module is a logit of the same size as that of the final output from the relevance network.

\subsubsection{Pretrain and Distill with Shifting Data}
From theorem \ref{theorem:xs}, we understand that the relevance network operates as a temporal data shift-free module.
Like the relevance network, the search-distill module is also used to approximate the probability $P(x_s|x)$, which can be trained using shifting data.
We first obtain the relevance network using the teacher network and loss function~(\ref{loss:retrieval_pretrain}).
Then, we employ \( \mathcal{D}_{\text{shifting}} \) and loss function~(\ref{loss:kd}) to perform distillation on the final logit layer of the search-distill module. The distillation process can be observed in Figure~\ref{fig:distillprocess}. 
We employ the mean squared error loss function to align the final logit layer of the search-distill module with that of the relevance network.

\vspace{-20pt}

\subsubsection{Finetune with Unshifting Data}
After obtaining a distilled search-distill module, we integrate the original model, the input $x_z$, and the search-distill module using a multilayer neural network for aggregation. 
This network is then trained with unshifting data and loss function~(\ref{loss:distill_finetune}), allowing us to finetune and enhance the original model with insights gleaned from shifting data without significantly increasing the deployment overhead of the original model.

\vspace{-10pt}

\section{Experiments}\label{sec:experiments}
In this section, we evaluate the RAD paradigm across multiple datasets using various traditional CTR models. 
We measure the efficacy of the RAD paradigm using the conventional CTR prediction task. We select two sets of classic CTR models: one representing factorization models and the other representing sequence prediction models. Each set of models is tested on three real-world datasets.

This section starts with five research questions (RQs) guiding the subsequent discourse.

\begin{itemize}
\item [\textbf{RQ1}] Does the RAD paradigm enhance the performance of existing CTR prediction models?
\item [\textbf{RQ2}] Can distillation functions be applied to the distillation of non-parametric models?
\item [\textbf{RQ3}] Can the relevance network, trained using shifting data, be transferred to predict future data?
\item [\textbf{RQ4}] How does the RAD paradigm affect the run time of the original model?
\item [\textbf{RQ5}] How do different distillation loss functions affect the RAD paradigm?
\end{itemize}

\subsection{Datasets}
In order to evaluate the performance of RAD, we conduct extensive experiments for CTR prediction tasks on three real-world large-scale recommendation datasets from Alibaba and Ant Group, i.e., Taobao, Tmall and Alipay\footnote{https://tianchi.aliyun.com/dataset/x, where `x' is `649', `42', and `53' for Taobao, Tmall, and Alipay, respectively.}. These datasets consist of sequential user-item interactions in e-commerce scenes.
The datasets contain timestamped user behaviors. We split them chronologically: the newest data for testing, the oldest for retrieval (\(\mathcal{D}_{\text{shifting}}\)), and the middle segment for training (\(\mathcal{D}_{\text{training}}\)). This mimics real-world scenarios where models must adapt to evolving preferences while leveraging historical data.





\subsection{Evaluation Metrics}
In the CTR prediction task, the area under the ROC curve (AUC) and negative log-likelihood (LogLoss) are widely used metrics. AUC measures the probability that a randomly chosen positive sample will be ranked higher than a randomly chosen negative sample, while Logloss quantifies the difference between the predicted probability and the actual label. We choose these two metrics to evaluate the model performance. 

\subsection{Compared Methods}



To evaluate our framework, we select six strong baseline models grouped into two categories: (1) \textbf{Feature interaction models}: DNN (MLP), DeepFM \cite{guo2017deepfm} (combines FM and DNN), and PNN \cite{qu2018product} (product-based feature interaction); (2) \textbf{Sequential models}: Transformer encoder \cite{vaswani2017attention}, DIN \cite{zhou2018deep}, and DIEN \cite{zhou2019deep} (DIN with RNN), which model user behavior sequences via attention mechanisms.

\begin{table*}[t]
\resizebox{0.95\textwidth}{!}{

\begin{tabular}{c|cccccccccc}
\toprule

\multicolumn{2}{c}{\multirow{2}{*}{Models}} & \multicolumn{3}{c}{Taobao} & \multicolumn{3}{c}{Tmall} & \multicolumn{3}{c}{Alipay}   \\ \cline{3-11} 
\multicolumn{2}{c}{} & LogLoss & AUC & Rel. Impr. & LogLoss & AUC & Rel. Impr.& LogLoss & AUC & Rel. Impr. \\
\hline
\multirow{6}{*}{Original} & DNN & 0.6454 & 0.6744 & - & 0.4259 & 0.8890 & - & 0.6230 & 0.7011 & -\\
& DeepFM & 0.6591 & 0.6545 & - & 0.4264 & 0.8885 & - & 0.6375 & 0.6803 & -  \\
& PNN & 0.6469 & 0.6787 & - & 0.4274 & 0.8949 & - & 0.6253 & 0.7130 & -  \\
\cline{2-11} 
& Transformer & 0.6757 & 0.5990 & - & 0.4644 & 0.8609 & - & 0.6761 & 0.6523 & - \\
& DIN & 0.6267 & 0.7040 & - & 0.4301 & 0.8820 & - & 0.6257 & 0.7333 & - \\
& DIEN & 0.6328 & 0.6947 & - & 0.3930 & 0.9037 & - & 0.6295 & 0.7218 & - \\
\hline
\hline

\multirow{6}{*}{Retrieval} & DNN & 0.4525 & 0.8633 & 28.00\% & 0.2985 & 0.9446 & 6.26\% & 0.4897	& 0.8455 & 20.60\% \\
& DeepFM & 0.4566 & 0.8614 & 31.61\% & 0.3016 & 0.9443 & 6.27\% & 0.4694&	0.8583 & 26.16\% \\
&PNN & 0.4576 & 0.8606 & 26.81\% & 0.3054 & 0.9468 & 5.80\% & 0.4429	& 0.8748 & 22.68\% \\
\cline{2-11} 
& Transformer & 0.4614 & 0.8557 & 42.86\% & 0.3459 & 0.9249 & 7.44\% & 0.5606 & 0.7984 & 21.23\% \\
& DIN & 0.5558 & 0.7906 & 12.30\% & 0.3239 & 0.9354 & 6.05\% & 0.5626 & 0.7826 & 6.72\%  \\
& DIEN & 0.5733 & 0.7702 & 10.87\% & 0.3226 & 0.9371 & 3.10\% &   0.5810 & 0.7684 & 6.46\%\\

\hline
\hline

\multirow{6}{*}{Distill} &DNN & 0.7201 & 0.7224 & 7.11\% & 0.4100 & 0.8961 & 0.80\% & 0.5972 & 0.7423 & 5.88\% \\
&DeepFM & 0.6144 & 0.7192 & 9.88\% & 0.4108 & 0.8961 & 0.86\% & 0.5900 & 0.7532 & 10.72\% \\
& PNN & 0.6106 & 0.7269 & 7.10\% & 0.4061 & 0.9026 & 0.87\% & 0.5686 & 0.7713 & 8.17\% \\
\cline{2-11} 
& Transformer & 0.6263 & 0.6953 & 16.09\% & 0.4237 & 0.8940 & 3.85\% & 0.6227 & 0.7196 & 10.3\% \\
& DIN & 0.5956 & 0.7420 & 5.40\% & 0.3812 & 0.9152 & 3.76\% & 0.5927 & 0.7425 & 1.25\% \\
& DIEN & 0.6099 & 0.7176 & 3.29\% & 0.3690 & 0.9157 & 1.33\% & 0.6230 & 0.7562 & 4.77\% \\

\toprule
\end{tabular}
}
\caption{Performance comparison of CTR prediction task. Rel.Impr means relative
AUC improvement rate against each original model. Improvements are statistically significant with $p < 0.01$.}
\label{tab:mainperformance}
\vspace{-20pt}
\end{table*}

\begin{table*}[t]
\begin{tabular}{c|ccccccc}
\toprule
\multicolumn{2}{c}{\multirow{2}{*}{Models}} & \multicolumn{2}{c}{Taobao} & \multicolumn{2}{c}{Tmall} & \multicolumn{2}{c}{Alipay}   \\ \cline{3-8} 
\multicolumn{2}{c}{}                        & LogLoss & AUC            & LogLoss & AUC           & LogLoss & AUC             \\
\hline
\multirow{3}{*}{pretrained with $\mathcal{D}_{\text{train}}$} & DNN & 0.6210 & 0.7139 & 0.3373	& 0.9291 & 0.6227 & 0.7070 \\

& DeepFM & 0.6239 & 0.7109 & 0.3337&	0.9320 & 0.6300 & 0.6966 \\
& PNN & 0.6188 & 0.7226 & 0.3272	& 0.9405& 0.6063 & 0.7339 \\

\hline
\multirow{3}{*}{pretrained with $\mathcal{D}_{\text{shifting}}$} & DNN & 0.5224	&0.8163 & 0.3141 & 0.9398 & 0.5006 &	0.8396 \\
                                        & DeepFM & 0.5183	& 0.8169 & 0.3103 & 0.9405   & 0.4724 & 0.8564 \\
                                        & PNN & 0.5087 &	0.8256 & 0.3179 & 0.9438 &0.4446 &	0.8744 \\
\toprule
\end{tabular}
\caption{Performance comparison of pretrained relevance network in Retrieval Framework.}
\label{tab:Invariance}
\vspace{-20pt}
\end{table*}



\subsection{Overall Performance (RQ1)}
The results are listed in table~\ref{tab:mainperformance}. The results reveal the following insights.
i) In our experiments, sequence models, except the original Transformer model, outperformed factorization models. This indicates that increasing data volume is beneficial for recommendation systems.
ii) The Retrieval Framework utilizes shifting data as a search space and pretrains the relevance network with shifting data, significantly enhancing performance beyond the original model. This demonstrates the substantial benefit of shifting data for the original model.
iii) The Distill Framework employs the distilled relevance network to assist the original model. While predictive performance surpasses the original model, it falls short of models augmented by the Retrieval Framework, suggesting that the distilled relevance network retains valuable information, albeit with less incremental information than the retrieval model.
iv) The performance of feature interaction-based models within the Distill Framework has already surpassed that of the original sequence models. Moreover, as illustrated in Figure~\ref{fig:time_distill}, the distillation method can enhance model performance without significantly increasing the inference burden

\subsection{Applying Knowledge Distillation to Non-Parametric Models (RQ2)}
Knowledge distillation is traditionally employed for transferring knowledge between parametric models, typically from a complex, larger model to a simpler, smaller one. 
In this context, we extend the application of knowledge distillation to transfer knowledge from a non-parametric model (BM25 search) to a simple neural network. 
As illustrated in Table~\ref{tab:mainperformance}, distilling information from the retrieval-based relevance network into the search-distill module demonstrates that the distilled student model can still significantly enhance the performance of the original model. However, compared to the direct use of the non-parametric model, performance is considerably reduced.
There is significant room for improvement in distilling information from non-parametric models into parametric models.

\vspace{-15pt}

\subsection{Temporal Invariance of Association (RQ3)}

Table~\ref{tab:Invariance} showcases a group of experiments designed to validate the theorem ~\ref{theorem:xs} of Temporal Invariance of Association. 
The relevance network aids the original model in prediction in two ways: the additional information provided by the retrieved similar data $x_s$, and the information gain brought by pretraining with shifting data. 
To eliminate the influence of $x_s$ on the results during testing, we added $D_{\text{train}}$ pretraining for comparison. 
Similarly, to rule out the impact of finetuning on the relevance network, we fixed the parameters of the relevance network during the finetuning phase.
By comparing the data in the table, we find that: i) The relevance network pretrained with shifting data performs better than the model pretrained only with train data. 
This indicates that the relevance network trained with shifting data indeed learns time-invariant information in the time-varying data distribution. 
Theorem \ref{theorem:xs} is validated. 
ii) The relevance network trained with $\mathcal{D}_{\text{shifting}}$ performs better because whether using $\mathcal{D}_{\text{shifting}}$ or $\mathcal{D}_{\text{train}}$, both are fitting $P(x_s|x)$, and shifting data has a much larger volume. Hence, the trained relevance network performs better. 
Thus, we obtain a network that enhances model training effectiveness by increasing the volume of data, unlike the network in figure~\ref{fig:tmallshift}, where increasing the volume of data would decrease model performance.
\vspace{-10pt}




\subsection{Performance Efficiency Study (RQ4)}

\begin{figure}[htbp]
    \centering
    \includegraphics[width=0.7\linewidth]{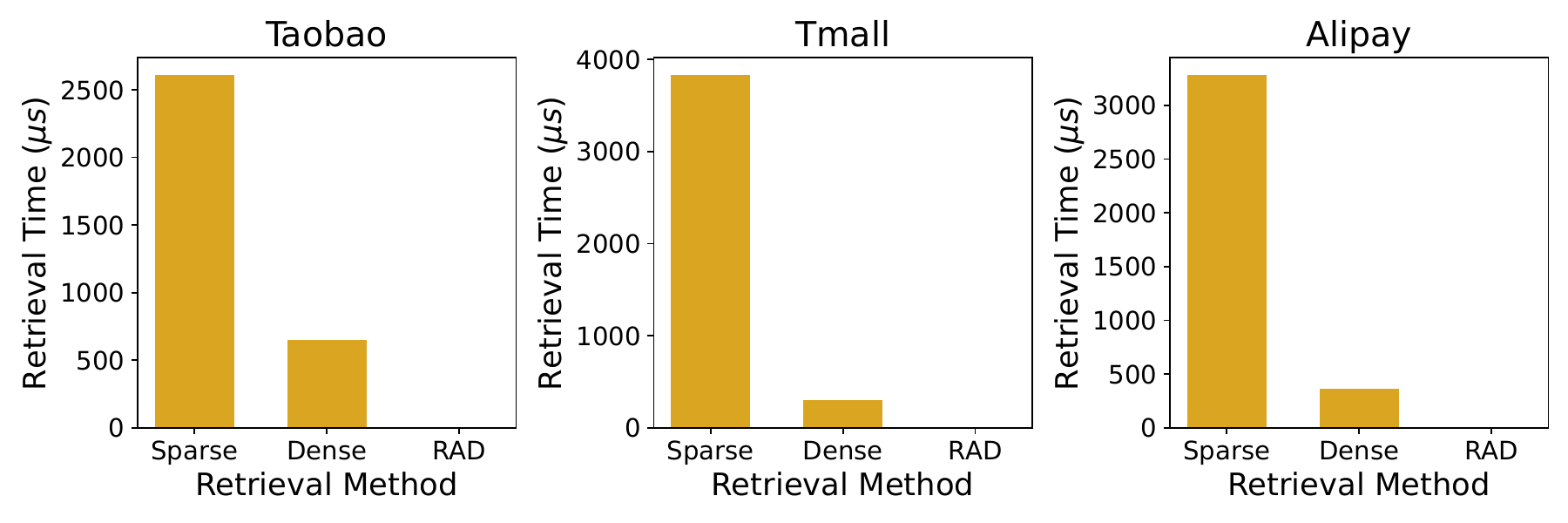}
\vspace{-10pt}
\caption{Comparison of Retrieval time.}
\vspace{-25pt}
\label{fig:time_retrieval}
\end{figure}

\begin{figure}[htbp]
    \centering
    \includegraphics[width=0.7\linewidth]{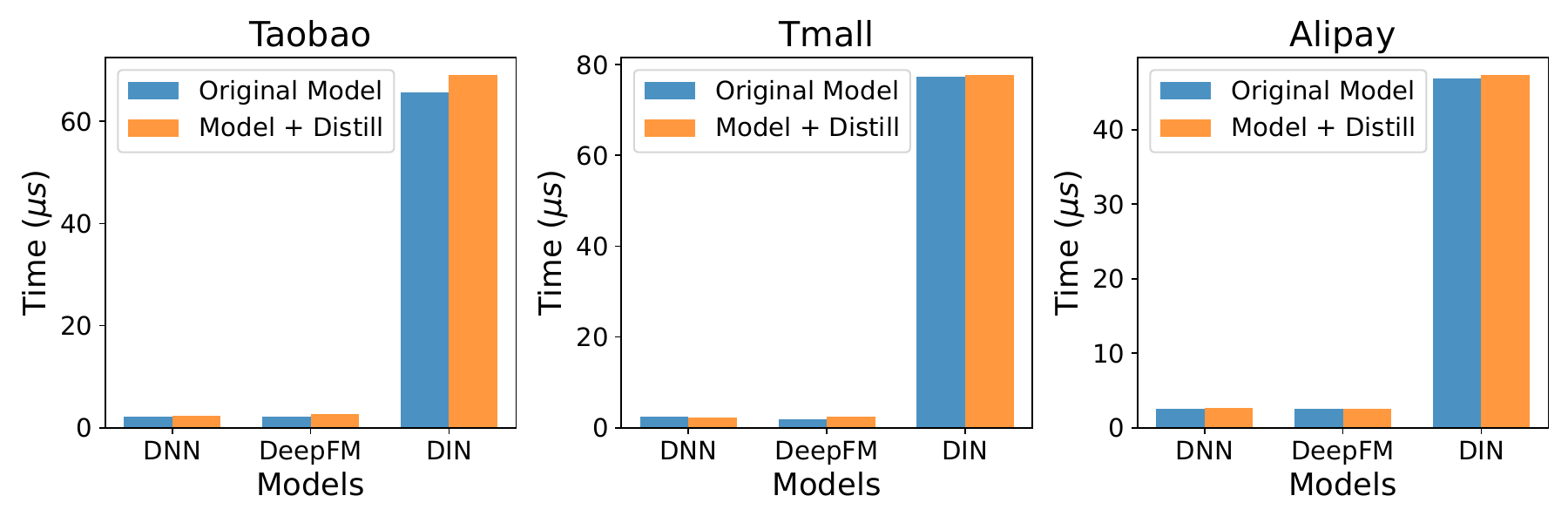}
\vspace{-10pt}
\caption{Comparison of Inference time.}
\vspace{-20pt}
\label{fig:time_distill}
\end{figure}

Recently, a range of retrieval-based models have been introduced in the field of recommender systems, some of which utilize discrete features for retrieval, such as SIM \cite{pi2019practice} and RIM\cite{qin2021retrieval}, while others employ continuous features, like DERT\cite{zheng2023dense}. Due to the significant time overhead associated with retrieval, these models face challenges in being directly deployed online. In Figure~\ref{fig:time_retrieval}, we compare three methods for augmenting relevant information. As RAD employs distillation to bypass the retrieval process, it incurs no retrieval time. On the other hand, directly aggregating the search-distill module with the original model increases the inference burden on the original model. However, as shown in Figure~\ref{fig:time_distill}, the additional inference load introduced by incorporating the search-distill module is not significant, and the time overhead it introduces is considerably less than the burden added by using sequence models for model inference.

\section{Related Work}\label{sec:related-work}
\noindent\textbf{Click-through rate (CTR) prediction}
is crucial in the fields of online advertising, recommendation systems, and information retrieval.
Many deep learning models for CTR prediction primarily concentrate on investigating the feature interactions within a single instance \cite{guo2017deepfm,he2017neural,qu2018product,liu2020autofis}. 
The Wide\&Deep network \cite{cheng2016wide} constructs fully connected (FC) layers based on field feature embeddings to extract feature interactions implicitly. 
Recently, graph neural networks have also shown great potential in recommendation scenes by constructing data into graph structures and performing message passing \cite{FIGNN}\cite{wang2019neural}\cite{he2020lightgcn}\cite{wang2019kgat}\cite{wu2021self}.
Despite the numerous models proposed in this direction, the model capacity might be constrained as the model's input is a single instance, necessitating the model to encode all knowledge into the parameters. 
Our work employs these deep models as the base model of the framework.

\vspace{5pt}

\noindent\textbf{Retrieval-based CTR prediction} methods, adapted from sequential recommendation, often replace RNNs when processing long user behavior sequences, as RNNs struggle with early pattern recall \cite{pi2019practice}, offering a more efficient alternative to recurrence and attention mechanisms.
RIM \cite{qin2021retrieval} expands the research scenario from sequential recommendation to generic tabular data prediction. The query is the target row data, and the retrieved data instances are the relevant rows in the table, with BM25 \cite{robertson1995okapi} serving as the predefined relevance function between two rows. 
DERT \cite{zheng2023dense} is the pioneering work to incorporate dense retrieval into recommendation systems. Retrieval-enhanced machine learning is now considered a significant frontier in machine learning research \cite{zamani2022retrieval}.

\vspace{5pt}\noindent
\textbf{Knowledge Distillation} (KD)
\cite{hinton2015distilling} seeks to transfer the 'dark knowledge' from a teacher model to a student model and is extensively used in model compression and knowledge transfer \cite{gou2021knowledge}. Recent studies have also employed KD in various recommendation tasks, such as click-through rate (CTR) prediction \cite{zhu2020ensembled}, topology distillation \cite{kang2021topology}, and multi-task transfer \cite{yang2022cross}.
Prior theoretical analyses \cite{tang2020understanding,phuong2019towards,chandrasegaran2022revisiting} have indicated that the teacher's predictions, also referred to as soft labels, are typically viewed as more informative training signals compared to binary hard labels. KD can adaptively adjust the training dynamics of the student model based on the values of the soft labels. Knowledge distillation is typically employed to transfer knowledge from a larger parameterized model to a smaller one. Recent studies have applied knowledge distillation to transfer knowledge from KNN to parameterized models \cite{zhang2023decoupled,yang2022nearest}, demonstrating the potential of knowledge distillation in non-parametric models.

\vspace{-10pt}

\section{Conclusion}

\vspace{-10pt}

In current recommendation systems, temporal data shift poses a significant challenge.
The presence of data shift prevents the system from simply enhancing the CTR model's adaptability to new data by adding more training data.
We observed that although the correlation between features and labels in recommendation systems changes over time, if a fixed search space is established, the relationship between the data and the search space remains invariant.
Therefore, we designed a framework that uses retrieval techniques to leverage shifting data for training a relevance network.
However, due to the use of BM25 as a retrieval method, this framework is challenging to deploy in online recommendation systems. 
We then designed a distillation method using knowledge distillation to transfer knowledge from the relevance network to a parameterized module, the search-distill module. We refer to this entire process as the Retrieval and Distill paradigm (RAD).
With the RAD paradigm, we have an effective method for leveraging shifting data to enhance the performance of CTR models.
In future research directions, we aim to incorporate a wider variety of data into the CTR model using RAD. On the other hand, enhancing the performance of the distillation method is also a significant area of focus.

\bibliographystyle{splncs04}
\bibliography{main}
%




\end{document}